\documentclass[aps,twocolumn,groupedaddress]{revtex4}
\usepackage{amsmath}
\usepackage{graphicx}
\begin{document}
\bibliographystyle{apsrev}
\title{Asymmetry of Nonlinear Transport and Electron Interactions in Quantum
Dots}
\author{D. M. Zumb\"uhl}
\affiliation{Department of Physics, Harvard University, Cambridge, Massachusetts
02138}
\author{C. M. Marcus}
\affiliation{Department of Physics, Harvard University, Cambridge, Massachusetts
02138}
\author{M. P. Hanson}
\affiliation{Materials Department, University of California, Santa Barbara,
California 93106}
\author{A. C. Gossard}
\affiliation{Materials Department, University of California, Santa Barbara,
California 93106}
\date{\today}
\begin{abstract}  The symmetry properties of transport beyond the linear regime
in chaotic quantum dots are investigated experimentally. A component
of differential conductance that is antisymmetric in both applied
source-drain bias $V$ and magnetic field $B$, absent in linear
transport, is found to exhibit mesoscopic fluctuations around a zero
average. Typical values of this component allow a measurement of the
electron interaction strength.
\end{abstract}
\maketitle

Quantum transport in disordered mesoscopic conductors and chaotic quantum dots
has been widely studied in the regime of linear response, and is well
understood in terms of universal statistical theories \cite{BeenakkerRMP}, even
in the presence of significant electron interaction \cite{ABGreview,
Alhassid}. A central principle of linear mesoscopic transport concerns the
symmetry of magnetoconductance: as a consequence of time reversal symmetry and
microscopic reversibility close to equilibrium, the differential conductance $g
= dI/dV$ of a two-terminal sample is symmetric in magnetic field $B$:
$g(B)=g(-B)$ \cite{Onsager, Casimir, Buttiker}, with generalized reciprocity
(Landauer-B\"uttiker) relations for multi-terminal coherent conductors
\cite{Buttiker}. Beyond linear response, i.e., at sufficient applied bias that
the current I is no longer proportional to source-drain voltage $V$, these
symmetry relations break down. Unless disallowed by some special symmetry,
differential conductance beyond linear response can generally contain a
component (here denoted $\tilde g$) that is proportional to both $B$ and $V$:
\begin{equation}
\label{Eq1} \tilde g=\alpha V B.
\end{equation}
For instance, contributions to $g$ of the form in Eq.~1 are permitted in
non-centrosymmetric materials \cite{Sturman}, chiral conductors
\cite{RikkenChiral}, including carbon nanotubes \cite{Ivchenko}, or conductors
with crossed electric and magnetic fields \cite{RikkenRelativistic}.

In the absence of electron interaction, including indirect interaction
such as inelastic phonon scattering, the coefficient $\alpha$ in Eq.~1
vanishes, since conduction at each energy within the finite bias window
independently obeys the symmetry of linear response and different energies do
not mix \cite{Larkin, Falko}. Moreover, as discussed recently in Refs.\
\cite{Sanchez, Spivak}, the coefficient $\alpha$ is proportional to the
electron interaction strength. This suggests that terms of the form of
Eq.~\ref{Eq1} can be used to measure interaction strength. As pointed out in
Refs.~\cite{Sanchez, Spivak}, finite $\tilde g$ arises from quantum
interference and is suppressed by decoherence and thermal averaging.
\begin{figure}[b]
\includegraphics{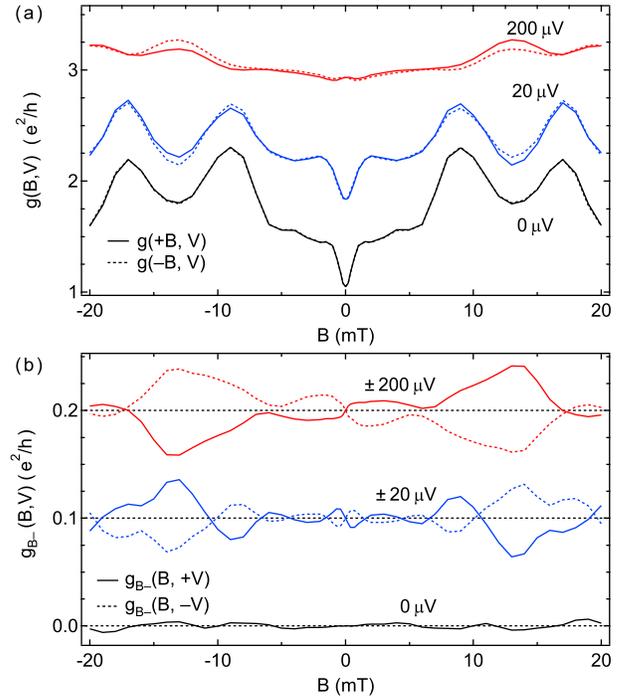} 
       \caption{\footnotesize {\label{fig1} (a) Differential
       conductance $g$ for $N=2$ as a function of magnetic field $B$ (solid
curves)
       and $-B$ (dashed curves) at source-drain bias $V$ as indicated, at base
       electron temperature $T_{el}=45\, \mathrm{mK}$. The blue (red)
       curves are offset by $\mathrm{+0.5\,(+1)\,e^2/h}$, respectively.
       (b) Antisymmetric in $B$ conductance $g_{B-}$ of the traces
       shown in (a), as a function of $B$ at $V$ (solid curves) and
       $-V$ (dashed curves) as indicated. The blue (red) curves are
       offset by $\mathrm{+0.1\,(+0.2)\,e^2/h}$, respectively. Black
       curves demonstrate $g(B,V=0)=g(-B,V=0)$ and blue/red curves
       indicate $g_{B-}$ is largely antisymmetric in $V$. }}
\end{figure}

This Letter presents a detailed study of the symmetry of nonlinear
conductance in an open chaotic GaAs quantum dot, using gate-controlled shape
distortion to gather ensemble statistics. We focus
particular attention on the component of $g$ that is odd in both $B$ and $V$, as
in Eq.~\ref{Eq1}, at moderately small source-drain voltage,
 $V\lesssim\Delta/e$, and fields, $B\lesssim \phi_0/A$, applied perpendicular
to the plane of the dot, where $\Delta$ is the average quantum level spacing in
the dot, $\phi_0=h/e$ is the flux quantum, and $A$ is the dot area. We find
that the component of differential conductance antisymmetric in $B$, denoted
$g_{B-}$, is also largely antisymmetric in $V$ and shows mesoscopic
fluctuations as a function of $B$, $V$, and shape-defining gate voltage $V_G$. As
anticipated theoretically \cite{Sanchez, Spivak}, we find that the average
coefficient $\alpha$ measured over an ensemble of dot shapes vanishes. The
standard deviation of $\alpha$, denoted $\delta \alpha$, does not vanish and is
used to characterize the strength of the interactions, as discussed below. The
dependence of $\delta \alpha$ on the number of modes $N$ in the
quantum-point-contact leads is found to be in disagreement with theory
\cite{Sanchez, Spivak}. However, present theory assumes that electrons do not
thermalize within the dot and that decoherence effects are negligible.
Accounting for an increased amount of thermalization within the dot at smaller
$N$ (consistent with independent measurements of electron distribution
functions \cite{ZumbuhlDistrib}) appears able to explain qualitatively the
observed dependence of $\delta \alpha$ on $N$.

\begin{figure}[t]
\includegraphics{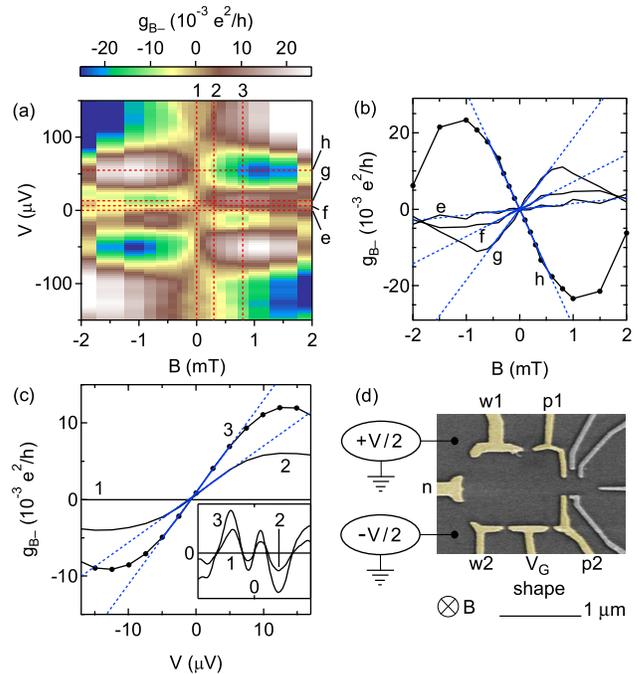}
\caption{\footnotesize{\label{fig2}(a) Antisymmetric component $g_{B-}$ (color
scale) as a
       function of $B$ and $V$ at $N=2$.
       Dashed lines indicate cuts
       shown in (b) and (c).
       (b) $g_{B-}$ as a function of $B$ at $V=0,7.5,15,55\,
           \mathrm{\mu V}$.
       (c) $g_{B-}$ as a function of $V$ at $B=0, 0.3, 0.8\,
           \mathrm{mT}$.
       Linear fits to curves in (b) and (c) are shown in dashed blue
       lines, with solid blue lines indicating the fitting range.
       Inset to (c) shows the same cuts as in (c) with the $V$
       axis extended to $\mathrm{\pm 125\,\mu V}$ and the $g_{B-}$
       axis extended to $\mathrm{\pm 25 \times 10^{-3} e^2/h}$.
       (d) Electron micrograph of a device with the same gate design as the
one measured.
Only the yellow gates
       are used in this experiment. $V$ is applied symmetrically. }}
\end{figure}

Measurements were carried out using a quantum dot of area $A \sim 1\,
\mathrm{\mu m^2}$ formed by Ti/Au depletion gates [see Fig.~\ref{fig2}(d)] on
the surface of a $\mathrm{GaAs/Al_{0.3}Ga_{0.7}As}$ heterostructure
$\mathrm{105\, nm}$ above the 2D electron gas. A bulk electron density
$n\sim\mathrm{2\times 10^{11}\, cm^{-2}}$ and mobility $\mu\sim\mathrm{2\times
10^5\, cm^2/Vs}$, giving a mean free path $\ell\sim1.5\, \mathrm{\mu m}$,
indicates ballistic transport within the dot. This device contains
$N_{dot}\sim2000$ electrons and has an average level spacing
$\Delta=2\pi\hbar^2/m^*A\sim 7\, \mathrm{\mu eV}$, where $m^*=0.067\, m_e$ is
the effective electron mass. The dot was designed to lack spatial symmetry
\cite{Lofgren} and is found to show universal statistics in linear conduction
characteristic of chaotic classical dynamics \cite{BeenakkerRMP, ABGreview,
Alhassid}. The gate marked ``shape", with voltage $V_G$ applied, was used to
modify the dot boundary, providing a means of generating ensemble statistics
\cite{Chan}. Gates $n$, $w1$, and $w2$ were actively modulated depending on
$V_G$ to maintain an integer number of quantum modes $N$ in each lead
throughout the ensemble of shapes. Gates $p1$ and $p2$ were strongly depleted,
isolating the adjacent smaller dot, which was used only to measure the electron
temperature ($T_{el}\mathrm{=45\pm5\, mK}$ at base) using Coulomb blockade peak
width.

To measure $g$, simultaneous lock-in measurements of both differential current
and voltage across the dot were made using a four-wire setup at 94.7 Hz with
typical ac excitation of $2\,\mu V$, in the presence of a dc source-drain
voltage $V$. All voltages were applied symmetrically across the dot
[Fig.~\ref{fig2}(a)] to reduce self-gating \cite{Lofgren}. The measurement
sequence was as follows: $g(B,V,V_G)$ was measured as a function of $V$
(innermost loop) and $B$ (2nd loop, with higher point-density around zero
field). Each of these two-parameter sweeps was repeated 20 times to reduce
noise (3rd loop), for each of 16 statistically independent shape gate voltages
$V_G$ (4th loop). Each multi-parameter sweep, which took about $30\, h$ to
complete, was measured at $N={1,2,4}$ modes in each lead (outermost loop).

As expected for a two terminal device \cite{Onsager, Casimir, Buttiker}, the
linear conductance $g(B, V=0)$ was found to be symmetric in $B$ within
measurement resolution, as seen in Fig.~\ref{fig1}(a). At finite bias, however,
the $\pm B$ symmetry is broken, as seen in the blue and red traces in
Fig.~\ref{fig1}(a). Decomposing $g$ into components that are symmetric
($g_{B+}$) and antisymmetric ($g_{B-}$) in field,
$g_{B\pm}(B,V)=\left(g(B,V)\pm g(- B,V)\right)/2$, we find $g_{B-}(B, V = 0)$
comparable to the measurement noise of $3\times10^{-3}e^2/h$ and barely
visible, while at $V = \pm20 \mathrm{\mu eV}$ and $ \pm200 \mathrm{\mu eV}$,
$g_{B-}$ is sizable and is found to be largely antisymmetric in $V$. The
typical amplitude of fluctuations of $g_{B-}$ is $\sim0.025\,e^2/h$, much
smaller than the amplitude of fluctuations of $g$, which is $\sim0.5\, e^2/h$.

The antisymmetry of $g_{B-}(B,V)$ in source-drain voltage $V$ is also seen in
Fig.~\ref{fig2}(a), which shows $g_{B-}(B,V)$ for $N=2$ as a two dimensional
color plot for a different shape sample than the one shown in Fig.~\ref{fig1}.
Note that $g_{B-}(B,V)$ changes sign repeatedly as a function of both $B$ and
$V$. Similar characteristics are seen for all dot shapes and values of $N$,
though for some shape samples the antisymmetry in $V$ is more pronounced than
in others.

\begin{figure}[t]
\includegraphics{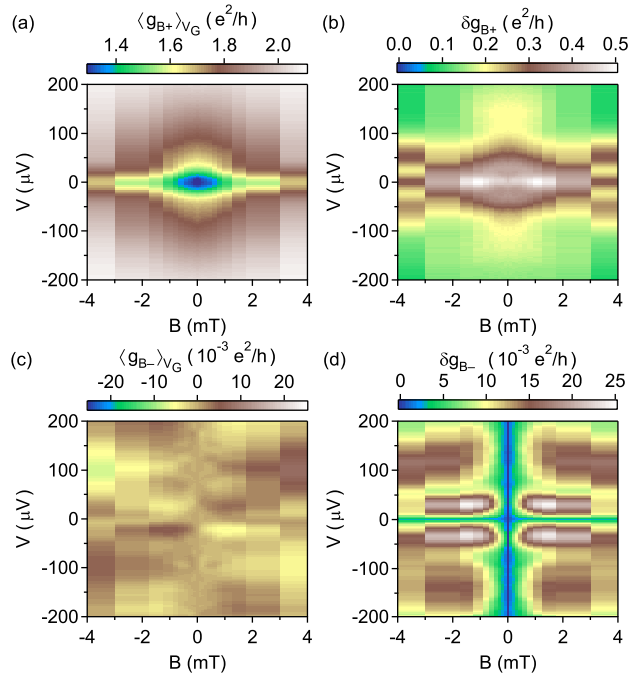}
\caption{\footnotesize{\label{fig3} Average
       conductance component $\langle g_{B+}\rangle_{V_G}$ (a) and
       $\langle g_{B-}\rangle_{V_G}$ (c) and standard deviation
       $\delta g_{B+}$ (b) and
       $\delta g_{B-}$ (d)
       (color scale) obtained from 16 independent shape samples at $N=2$ modes
as a function of $B$ and $V$.
       $\langle g_{B-} \rangle$ is
       zero within statistical uncertainty; standard deviations are used
       to characterize the antisymmetries.
       $\langle g_{B+}\rangle$ shows a strong dip at $(B,V)=0$ due
       to ballistic weak localization.
      }}
\end{figure}

\begin{figure}[t]
\includegraphics{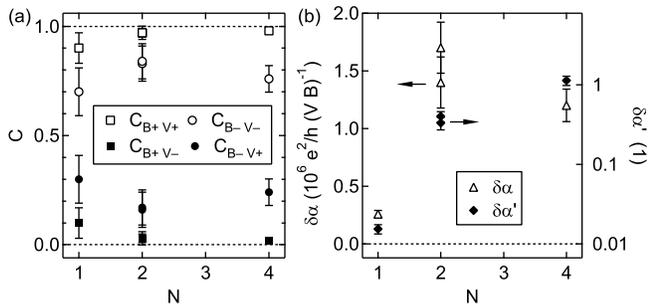}
\caption{\footnotesize{\label{fig4} (a) Normalized symmetry
       parameter $C$ as a function of the number of modes $N$
       for all $B$ and $V$ symmetries as indicated,
       giving a statistical characterization of the nonlinear
       field symmetry properties of conductance over all $V$, $B$ and shape
       samples measured.
       (b) Standard deviation $\delta\alpha$ (open triangles)
       and dimensionless standard deviation $\delta \alpha'$ (solid diamonds)
       of the interaction parameter (see text) as a function of $N$.
      }}
\end{figure}

For small fields,  $B\lesssim \phi_0 /A$,  $g_{B-}(B)$ is proportional to $B$
and becomes sizable for $V\gtrsim\Delta/e\sim7\, \mathrm{\mu eV}$, as seen in
Fig.~\ref{fig2}(b). At larger fields, $B\gtrsim \phi_0 /A$, $g_{B-}(B)$ shows
mesoscopic fluctuations as a function of $B$, including sign changes. The field
scale where this crossover occurs is consistent with the field scale of weak
localization [see Fig.~\ref{fig3}(a)] and conductance fluctuations in this
device (not shown). This field scale is somewhat smaller than $\phi_0 /A \sim4
\mathrm{mT}$, similar to previous experiments \cite{Marcus, ZumbuhlCF} because
the relevant area is not the dot area but the area of typical trajectories
before escape, which is larger than A by a factor $\sim \sqrt[4]{N_{dot}}$
\cite{CMchaosReview}.
For $V\lesssim\Delta/e$, $g_{B-}(V)$ is also found to be proportional to $V$ and
becomes sizable for $B\gtrsim1\, \mathrm{mT}$, as seen in Fig.~\ref{fig2}(c).
For $V\gtrsim\Delta/e$, $g_{B-}(V)$ starts to deviate from the linear
dependence on $V$ and shows mesoscopic fluctuations as a function of $V$, also
including sign changes. As mentioned above, while some shapes show more
pronounced antisymmetry of $g_{B-}$ with respect to $V$ than others, $g_{B-}$
is predominantly antisymmetric in $V$, see discussion of Fig.~\ref{fig4}(a).
The characteristic scales $V\sim\Delta/e$ and $B\sim \phi_0 /A$ for mesoscopic
fluctuations of $g_{B-}(B,V)$ provide assurance that this component arises from
coherent transport within the dot.

The average symmetric component of conductance, $\langle g_{B+}\rangle_{V_G}$,
which is maximal around $V = 0$, has a pronounced minimum around $B=0$ due to
ballistic weak localization \cite{Huibers}, as seen in Fig.~\ref{fig3}(a). From
the magnitude of the weak localization feature in $\langle
g(B,V=0)\rangle_{V_G}$, we extract a base-temperature phase coherence time
$\tau_\varphi\sim2\, \mathrm{ns}$, consistent with previous experiments
\cite{Huibers}. As seen in Fig.~\ref{fig3}(a), finite $V$ reduces the dip
$\langle g(B,V=0)\rangle_{V_G}$ around $B= 0$, presumably the result of
dephasing caused by heating and non-equilibrium effects \cite{Switkes, Linke}.
Figure~\ref{fig3}(b) shows that the standard deviation of the symmetric
component $\delta g_{B+}$ is peaked around $B=0$, a well-known effect
associated with the breaking of time-reversal symmetry \cite{Chan,
BeenakkerRMP,ABGreview, Alhassid}. Similar to the weak localization correction,
$\delta g_{B+}$ is also reduced at finite bias. In this case, the reduction is
due both to dephasing and an explicit dependence on temperature and energy
averaging.

Figure 3(c) shows the shape-averaged antisymmetric component of conductance
$\langle g_{B-}\rangle_{V_G}$, which is zero within statistical uncertainty
(based on 16 shape samples), indicating that fluctuations of $g_{B-}$ cancel
each other
upon averaging, as predicted theoretically \cite{Sanchez, Spivak}. As seen in
Fig.~\ref{fig3}(d), the standard deviation $\delta g_{B-}$ becomes sizable on a
voltage scale of order of the dot level spacing, with a maximum at moderate
$|V|$, and then decreases for larger $|V|$, presumably due to heating or
decoherence effects. By definition, $\delta g_{B-}$ is zero for $B=0$ and, as
expected for a $g_{B-}$ that is largely antisymmetric in $V$, $\delta g_{B-}$
is comparable to the statistical error along the $V=0$ line.

To characterize symmetry in terms of statistical quantities, we define
normalized symmetry parameters $C_{B\pm V\pm}=\delta^2 g_{B\pm V\pm}/\delta^2
g_{B\pm}$, where $\delta^2 g_{B\pm V\pm}$ is the variance of the component of
$g$ with a particular symmetry with respect to $B$ and $V$. Here, ensemble
averaging is performed over the entire $(B,V,V_G)$ parameter space, resulting
in approximately $5\times4\times16$ statistically independent samples. We note
that $C$ ranges from zero
to one, and $C_{B\pm V\pm}+C_{B \pm V \mp}=1$. As seen in Fig.~\ref{fig4},
$C_{B+V+}\sim 1$ (open squares) and $C_{B+V-}\sim 0$ (filled squares) within
the statistical uncertainty, showing that $g_{B+}$ is symmetric in $V$ for all
measured conductance samples and modes numbers $N$. $C_{B-V-}\sim 0.75$ (open
circles) and $C_{B-V+}\sim 0.25$ (filled circles) without a strong $N$
dependence, indicating that a significant part of $g_{B-}$ is antisymmetric in
$V$, though the antisymmetry is not perfect.

The interaction coefficient $\alpha$ was extracted for each shape by performing
fits linear in both $B$ and $V$ to $g_{B-}(B,V)$ for small $V < \Delta/e$ and
$B<\phi_0/A$, similar to the fits shown in Figs.~\ref{fig2}(b) and (c). The
resulting $\alpha$ is seen to have mesoscopic fluctuations and changes sign
frequently, consistent with $\langle \alpha \rangle_{V_G} = 0$ within statistical
uncertainty. The standard deviation $\delta \alpha$, used to characterize the
strength of electron interaction, is shown in Fig.~\ref{fig4}(b) (open triangles)
as a function of the mode number $N$.

For a comparison of the interaction parameter $\alpha$ to theory, we note that
Ref.~\cite{Spivak} predicts $\delta \alpha$ for $kT \ll N\Delta$, $V\ll
\Delta/e$ and $B\ll \phi_0/A$:
\begin{equation}
\label{Eq2}
\delta\alpha = \delta\alpha' \frac{1}{2N^2}
\frac{e}{\Delta}\frac{A}{\phi_0}\frac{e^2}{h}
\end{equation}
where $\delta\alpha'$ is a dimensionless parameter characterizing the electron
interaction strength.  Using a different approach, Ref.~\cite{Sanchez} arrives
at a similar expression for $\delta \alpha$, with the same dependencies on $N$
and $\Delta$. To facilitate a comparison of experiment and theory, we plot the
dimensionless quantity $\delta\alpha'$ in Fig.~\ref{fig4}(b), noting a
pronounced dependence of $\delta\alpha'$ on $N$. As defined in Eq.~2,
$\delta\alpha'$ is not expected to depend on $N$, according to
Refs.~\cite{Sanchez, Spivak}. A likely reconciliation is that for $N<4$, the
escape time from the dot is sufficiently long so that electrons have time to
equilibrate (thermalize at a higher temperature) before escaping. As a
result, $\delta \alpha'$ is reduced at $N=1,2$ compared to $N=4$. Indeed, from
direct measurement of electron distribution functions in the same device
\cite{ZumbuhlDistrib}, it is known that for $N=1$ the distribution of electron
energies in the dot is thermal, with an elevated effective temperature that
depends on $V$. For $N=4$, on the other hand, with a shorter electron dwell
time in the dot, nonthermal distributions are seen
\cite{ZumbuhlDistrib}. For the case $N=4$, the measured
$\delta\alpha'=1.1\pm0.2$ is reasonably consistent with the theoretical
estimate $\delta \alpha' = \pi$ \cite{Sanchez}, which assumes perfect screening
of electrons. Other factors possibly contributing to the observed $N$
dependence of $\delta\alpha'$ may include imperfectly transmitting modes or
mode mixing as well as finite temperature and/or decoherence effects, all of
which are not currently accounted for theoretically.

In conclusion, we have investigated the magnetic field asymmetry of conductance
beyond the linear regime in gate-defined quantum dots. The conductance
component $g_{B-}$, which is defined to be antisymmetric in $B$, was found to
be also predominantly antisymmetric in $V$ and is of the form of Eq.~\ref{Eq1}
for $V\ll\Delta/e$ and $B\ll\phi_0/A$. The interaction coefficient $\alpha$,
extracted from linear fits to $g_{B-}$, has mesoscopic fluctuations with zero
average. Comparison to recent theory \cite{Sanchez, Spivak} is most appropriate
for the data with four modes per lead, where electrons remain out of
equilibrium during the short dwell time. In this case, consistency with theory
\cite{Sanchez} relating $\alpha$ to the interaction strength appears
reasonable, suggesting good electron screening.

We thank Boris Spivak for suggesting this problem to us and Markus B\"uttiker
and David Sanchez for numerous contributions. This work was supported in part by
DARPA under QuIST, ARDA/ARO Quantum Computing Program, and Harvard NSEC. Work at
UCSB was supported by QUEST, an NSF Science and Technology Center.

\emph{Note added.} During completion of this manuscript, closely related
experimental work on carbon nanotubes appeared \cite{Cobden}.


{ \small 
 }
\end{document}